\documentclass{ifacconf}

\usepackage{graphicx}      
\usepackage{natbib}        
\usepackage{amssymb}  
\usepackage{amsmath}
\usepackage{color}
\usepackage{multicol}
\usepackage{enumerate}

\begin{document}
\begin{frontmatter}

\title{Discontinuous energy shaping control of the Chaplygin sleigh } 


\author[First]{Joel Ferguson}
\author[Second]{Alejandro Donaire}
\author[First]{Richard H. Middleton}

\address[First]{School of Electrical Engineering and Computing and PRC CDSC, The University of Newcastle, Callaghan, NSW 2308, Australia (e-mail: Joel.Ferguson@uon.edu.au; Richard.Middleton@newcastle.edu.au).}
\address[Second]{Department of Electrical Engineering and Information Theory and PRISMA Lab, University of Naples Federico II, Napoli 80125, Italy, and with the School of Electrical Eng. and Comp. Sc. of the Queensland University of Technology, Brisbane, QLD, Australia (e-mail: alejandro.donaire@qut.edu.au)}

\begin{abstract}                
In this paper we present an energy shaping control law for set-point regulation of the Chaplygin sleigh. It is well known that nonholonomic mechanical systems cannot be asymptotically stabilised using smooth control laws as they do no satisfy Brockett's necessary condition for smooth stabilisation. Here, we propose a discontinuous control law that can be seen as a potential energy shaping and damping injection controller. The proposed controller is shown to be robust against the parameters of both the inertia matrix and the damping structure of the open-loop system.
\end{abstract}


\end{frontmatter}

\section{Introduction}
Mechanical systems are often subject to constraints which restrict the motion of the system. These constraints are often categorised as begin either \emph{holonomic} or \emph{nonholonomic}. Holonomic constraints refer to static relationships between configuration variables which, in effect, restricts the configuration space of a system. Nonholonomic constraints, however, refers to all constraints that cannot be described in this manner \citep{goldstein1965classical}. Of particular interest to this work, the constraints that arise from non-slip condition of wheels are necessarily described as a relationship between the configuration and velocity of a system \citep{Bloch2003}. As such, these constraints restrict in what directions the system can move and therefore, they are nonholonomic. In this work, we consider the Chaplygin sleigh which is a benchmark system widely used for nonholonomic control design \citep{VanDerSchaft1994,Astolfi1996,Lee2007,Fujimoto2012,Tian2002,Bloch1992}.

Nonholonomic systems with constraints that are linear in velocities can be represented as port-Hamiltonian (pH) system with Lagrange multipliers that enforce the constraints. In the work of \cite{VanDerSchaft1994} it was shown that by reducing the dimension of the momentum space, these systems have an equivalent representation without Lagrange multipliers. The reduced representation is essentially `constraint free' insofar as any state in the reduced state-space is permissible. Important to this work, the Chaplygin sleigh admits such a representation \citep{Astolfi2010}.

Here, we utilise this `constraint free' representation of the Chaplygin system to develop a control law to achieve set-point regulation of the system. Unfortunately, as it is well known, Brockett's necessary condition for asymptotic stabilisation using smooth feedback control is not satisfied by nonholonomic mechanical systems. As a consequence, this class of system cannot be stabilised using continuously differentiable control laws \citep{Brockett1983}. This restriction does not rule out the possibility of asymptotic stabilisation using non-smooth controllers, which has been achieved in \citep{Astolfi1996,Fujimoto2012}. In this work, we propose a discontinuous energy shaping control law for the Chaplygin system.

While control of the Chaplygin system (and nonholonomic systems generally) has been extensively studied, control methods that exploit the natural passivity of the system are quite limited. Similar to the method proposed here, a discontinuous energy shaping control law was proposed by \cite{Fujimoto2012} for the rolling coin system---which is encompassed in the Chaplygin system used here---to asymptotically stabilise the system. A different approach was taken by \cite{Lee2007} where a switching strategy was used to drive a mobile robot---which again is encompassed in the Chaplygin system used here---to a compact set containing the origin. 

Previously, we studied the control of the Chaplygin system in \citep{Ferguson2016} by switching between two manifold regulating control laws where each law could be considered to be energy shaping controllers. Here, we extend this previous work by proposing a single energy shaping control law that drives the configuration of the system to the origin. By exploiting the passivity properties of the open-loop system, the controller is robust against both the inertia and damping matrices.

The remained of the paper is structured as follows: The Chaplygin sleigh model is presented and the problem formulated in section \ref{probForm}. In Section \ref{control} the discontinuous, potential energy shaping controller is presented and the stability properties of the closed-loop are analysed in section \ref{stab}. A numerical simulation of the closed-loop is presented in Section \ref{simulation} and conclusions drawn in section \ref{conclusion}.

\textbf{Notation:}
For a differentiable function $\mathcal{H}(x)$, $\nabla\mathcal{H}$ denotes the column vector of partial derivatives $\frac{\partial^\top \mathcal{H}}{\partial x}$. Given a differentiable function $f(x)\in\mathbb{R}^n$, $\frac{\partial f}{\partial x}$ denotes the standard Jacobian matrix. $0_n$ is a matrix of dimension $n\times n$ with all elements equal to zero whereas $0_{n\times m}$ is a $n\times m$ matrix of all zeros. $I_n$ denotes a $n\times n$ dimension identity matrix.

\section{Problem formulation}\label{probForm}
\subsection{Chaplygin sleigh model}
\begin{figure}
	\centering
	\includegraphics[trim = 5mm 3mm 5mm 3mm, clip, width=.7\linewidth]{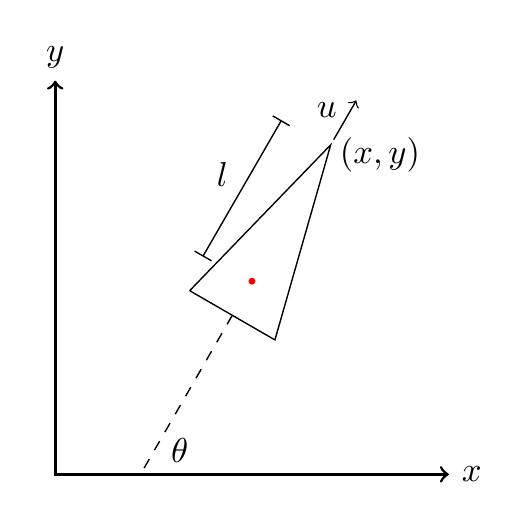}
	\caption{The Chaplygin sleigh is fixed to the ground at the point $(x,y)$. It is able to pivot about this point and move forwards in the $u$ direction. The point $(x,y)$ is constrained from moving in a direction perpendicular to $u$. The centre of mass is indicated by the red spot and is a distance $l$ from the point $(x,y)$.}
	\label{system:chaplygin}
\end{figure}

This paper is focused on control design for the Chaplygin sleigh system (Figure \ref{system:chaplygin}). This system can be modelled as a pH system of the form \citep{Astolfi2010}:
\begin{equation}\label{probForm:ChaplyginOL}
	\begin{split}
		\begin{bmatrix}
			\dot{q} \\ \dot{p}
		\end{bmatrix}
		&=
		\begin{bmatrix}
			0_3 & Q(q) \\
			-Q^\top(q) & J(p)-D(q,p)
		\end{bmatrix}
		\begin{bmatrix}
			\nabla_q\mathcal{H} \\ 
			\nabla_p\mathcal{H}
		\end{bmatrix}
		+
		\begin{bmatrix}
			0_{3\times 2} \\ 
			I_2 \\ 
		\end{bmatrix}
		u \\
		y
		&=
		\nabla_p\mathcal{H} \\
		\mathcal{H} 
		&= 
		\frac{1}{2}p^\top M^{-1}p,
	\end{split}
\end{equation}
with generalised coordinates $q = (x,y,\theta)$, where $x$ and $y$ denote the position at which the sleigh is fixed to the ground in the $x-y$ plane and $\theta$ describes the sleigh's heading angle, $p = (p_1,p_2)$ is the momentum and $D(q,p)$ is the damping matrix satisfying $D = D^\top\geq 0$.
The system matrices are given by
\begin{align}\label{system:modelq}
	Q(q)
	&=
	\begin{bmatrix}
		\cos{\theta} & 0 \\
		\sin{\theta} & 0 \\
		0 & 1
	\end{bmatrix} 
	\nonumber \\
	J(p)
	&=
	\begin{bmatrix}
		0 & \frac{ml}{J+ml^2}p_2 \\
		-\frac{ml}{J+ml^2}p_2 & 0
	\end{bmatrix}
	\nonumber \\
	M
	&=
	\begin{bmatrix}
		m & 0 \\
		0 & J + ml^2
	\end{bmatrix},
\end{align}
where $m$ is the mass of the sleigh and $J$ is the rotational inertia about the centre of mass. 

In the special case where $l=0$, this system coincides with the knife edge system \citep{Bloch2003}. The knife edge system is closely related to the rolling disk system, studied in \citep{Fujimoto2012}, which has an extra state associated with the roll angle of the disk. If the roll angle from the rolling disk is ignored, it coincides with the knife edge system. Due to the relationship between these systems, the regulating controller developed in this paper can be applied to either of these examples.

\subsection{Problem statement}
Considering the Chaplygin sleigh system \eqref{probForm:ChaplyginOL}, our objecting is to design a control law
\begin{equation}
	u = u(q,p)
\end{equation}
such that $\lim_{t\to\infty}q(t) = 0_n$.

\section{Discontinuous control law}\label{control}
The approach taken to control the Chaplygin system is to first perform two consecutive coordinate transformations, $q\to z\to w$. The control law is then designed in the $w$ coordinates where the control objective can be achieved by potential energy shaping using a quadratic potential function. The stability analysis relies heavily on the relationship between the $z$ and $w$ coordinates.

\subsection{Transformation $q\to z$}
The first of our coordinate transformations $q\to z$ is defined by the mapping
\begin{equation}\label{control:zTransform}
	z
	=
	\begin{bmatrix}
		z_1 \\ z_2 \\ z_3
	\end{bmatrix}
	=
	f_z(q)
	\triangleq
	\begin{bmatrix}
		0 & 0 & 1 \\
		\cos{\theta} & \sin{\theta} & 0 \\
		\sin{\theta} & -\cos{\theta} & 0
	\end{bmatrix}
	\begin{bmatrix}
		x \\ y \\ \theta
	\end{bmatrix}.
\end{equation}
The Chaplygin system \eqref{probForm:ChaplyginOL} can be expressed in the coordinates $(z,p)$ by the equations
\begin{equation}\label{ctrl:ChaplyginOLz}
	\begin{split}
		\begin{bmatrix}
			\dot{z} \\ \dot{p}
		\end{bmatrix}
		&=
		\begin{bmatrix}
			0_3 & Q_z(z) \\
			-Q_z^\top(z) & J(p)-D_z(z,p) \\
		\end{bmatrix}
		\begin{bmatrix}
			\nabla_z\mathcal{H} \\ 
			\nabla_p\mathcal{H}
		\end{bmatrix}
		+
		\begin{bmatrix}
			0_{3\times 2} \\ 
			I_2 \\ 
		\end{bmatrix}
		u \\
		y
		&=
		\nabla_p\mathcal{H} \\
		\mathcal{H} 
		&= 
		\frac{1}{2}p^\top M^{-1}p,
	\end{split}
\end{equation}
with $Q_z$ and $D_z$ defined by 
\begin{align}\label{ctrl:ChaplyginOLzMat}
	Q_z(z)
	&=
	\frac{\partial f_z}{\partial q}(q)Q(q)|_{q=f_z^{-1}(z)}
	=
	\begin{bmatrix}
		0 & 1 \\
		1 & -z_3 \\
		0 & z_2
	\end{bmatrix}
	\nonumber \\
	D_z(z,p)
	&=
	D(p,q)|_{q=f_z^{-1}(z)}.
\end{align}
The reason for expressing the system as a function of $z$ is the structure of $Q_z$ in \eqref{ctrl:ChaplyginOLzMat}. Importantly, $Q_z$ has a full-rank left annihilator
\begin{equation}\label{ctrl:QzPerp}
	Q_z^\perp(z)
	=
	\begin{bmatrix}
		-z_2 & 0 & 1
	\end{bmatrix}.
\end{equation}

\subsection{Transformation $z\to w$}
Similar to the transformation to $z$ in the previous subsection, we now consider the coordinate transformation $z\to w$. The proposed transformation is given by
\begin{equation}\label{control:wTransform}
	w
	=
	\begin{bmatrix}
		w_1 \\ w_2 \\ w_3
	\end{bmatrix}
	=
	f_w(z)
	\triangleq
	\begin{bmatrix}
		z_1 \\
		z_2z_1^{-1} - 2z_3z_1^{-2} \\
		2z_3z_1^{-2}
	\end{bmatrix}.
\end{equation}
The inverse transformation $z = f_w^{-1}(z)$ is given by
\begin{equation}\label{control:wTransformInv}
	z
	=
	\begin{bmatrix}
		z_1 \\ z_2 \\ z_3
	\end{bmatrix}
	=
	f_w^{-1}(w)
	\triangleq
	\begin{bmatrix}
		w_1 \\
		w_1w_2 + w_1w_3 \\
		\frac12 w_1^2w_3
	\end{bmatrix}.
\end{equation}
This are two important properties that have motivated this choice for $f_w$: Firstly, the mapping $f_w^{-1}:w\to z$ is smooth. This means that if a solution $w(t)$ is bounded, then $z(t)$ is be bounded also. Secondly, the entire set $\{w\in\mathbb{R}^3|w_1=0\}$ corresponds to $z=0_{3\times 1}$. This means that in the $w$ coordinates, the control objective can be addressed simply by regulating the variable $w_1$ to zero whilst keeping $w_2$ and $w_3$ bounded.

The Chaplygin system \eqref{ctrl:ChaplyginOLz} can be expressed as a function of $(w,p)$ by the equations
\begin{equation}\label{ctrl:ChaplyginOLw}
	\begin{split}
		\begin{bmatrix}
			\dot{w} \\ \dot{p}
		\end{bmatrix}
		&=
		\begin{bmatrix}
			0_3 & Q_w(w) \\
			-Q_w^\top(w) & J(p)-D_w(w,p) \\
		\end{bmatrix}
		\begin{bmatrix}
			\nabla_w\mathcal{H} \\ 
			\nabla_p\mathcal{H}
		\end{bmatrix}
		+
		\begin{bmatrix}
			0_{3\times 2} \\ 
			I_2 \\ 
		\end{bmatrix}
		u \\
		y
		&=
		\nabla_p\mathcal{H} \\
		\mathcal{H} 
		&= 
		\frac{1}{2}p^\top M^{-1}p,
	\end{split}
\end{equation}
with $Q_w$ and $D_w$ defined by
\begin{align}\label{ctrl:ChaplyginOLwMat}
	Q_w(w)
	&=
	\frac{\partial f_w}{\partial z}(z)Q_z(z)|_{z=f_w^{-1}(w)} \nonumber \\
	&=
	\begin{bmatrix}
		0 & 1 \\
		1 & -\frac{2w_2}{w_1}-w_3-\frac12 w_1^2w_3 \\
		0 & \frac{2w_2}{w_1^2}
	\end{bmatrix}
	\nonumber \\
	D_w(w,p)
	&=
	D_z(p,z)|_{z=f_w^{-1}(w)}.
\end{align}
Importantly, the matrix $Q_w$ is ill-defined at $w_1 = 0$. This has the consequence of the dynamics \eqref{ctrl:ChaplyginOLw} being undefined at this point. As such we define the set on which the dynamics \eqref{ctrl:ChaplyginOLw} are defined:
\begin{equation}
	U
	=
	\{
		(w,p)\in\mathbb{R}^5|w_1\neq 0
	\}.
\end{equation}
The dynamics \eqref{ctrl:ChaplyginOLw} are well defined on the set $U$.

\subsection{Regulation control law}
Consider the following control law as a solution to the problem statement:
\begin{equation}\label{ctrl:CtrlLaw}
	\begin{split}
		u(w,p)
		&=
		-
		Q_w^\top Lw
		-
		\hat D\nabla_p\mathcal{H}
		-
		\underbrace{
		\begin{bmatrix}
			0 & 0 \\
			0 & \frac{k}{w_1^2}
		\end{bmatrix}}_{D_i(w)}
		\nabla_p\mathcal{H}
	\end{split}
\end{equation}
where $\hat D\in\mathbb{R}^{2\times 2}$ is positive definite, $L=\operatorname{diag}(l_1,l_2,l_3)$ where each $l_i\in\mathbb{R}$ is positive and $k>0$ is a constant.

\begin{rem}
Considering the Chaplygin system \eqref{ctrl:ChaplyginOLz}, the term $\nabla_p\mathcal{H}$ can be expressed as a function of $z,\dot z$ as
\begin{equation}
	\nabla_p\mathcal{H}
	=
	\begin{bmatrix}
		z_3 & 1 \\ 1 & 0
	\end{bmatrix}
	\begin{bmatrix}
		\dot z_1 \\ \dot z_2
	\end{bmatrix}.
\end{equation}
Thus, the control law \eqref{ctrl:CtrlLaw} can be written independent of the mass matrix $M$.
\end{rem}
\begin{rem}
The control law \eqref{ctrl:CtrlLaw} is independent of the open-loop damping $D_w$ and is, thus, robust against this parameter.
\end{rem}
\begin{rem}
The control law \eqref{ctrl:CtrlLaw} has been given as a function of $(w,p)$ but can be equivalently expressed as a function of $(z,p)$ or $(q,p)$ using the mappings $f_w$ \eqref{control:wTransform} and $f_z$ \eqref{control:zTransform}.
\end{rem}

Now we show that the closed-loop system admits a Hamiltonian representation.
\begin{prop}
Consider the Chaplygin system \eqref{ctrl:ChaplyginOLw} in closed-loop with the control law \eqref{ctrl:CtrlLaw}. The closed-loop dynamics are given by
\begin{equation}\label{ctrl:ChaplyginCLw}
	\begin{split}
		\begin{bmatrix}
			\dot{w} \\ \dot{p}
		\end{bmatrix}
		&=
		\begin{bmatrix}
			0_{n\times n} & Q_w(w) \\
			-Q_w^\top(w) & J(p)-D_d(w,p) \\
		\end{bmatrix}
		\begin{bmatrix}
			\nabla_w\mathcal{H}_d \\ 
			\nabla_p\mathcal{H}_d
		\end{bmatrix} \\
		\mathcal{H}_d
		&= 
		\frac{1}{2}p^\top M^{-1}p + \frac12 w^\top Lw,
	\end{split}
\end{equation}
where
\begin{equation}
D_d(w,p)
=
D_w(w,p)+\hat D+D_i(w).
\end{equation}
\end{prop}
\begin{pf}
	The proof follows from direct matching. \qed
\end{pf}

The control law \eqref{ctrl:CtrlLaw} can be interpreted as potential energy shaping plus damping injection. To see this, first notice that the role of the term $-Q_w^\top Lw$ is to add the term $\frac12 w^\top Lw$ to the closed-loop Hamiltonian. This term can be considered a potential function in $w$. Secondly, the term $-\hat D\nabla_p\mathcal{H}-D_i\nabla_p\mathcal{H}$ is to increase the damping from $D_w$ to $D_d$ in closed-loop.

\section{Stability analysis}\label{stab}
We now analyse the asymptotic behaviour of the closed-loop system \eqref{ctrl:ChaplyginCLw}. Note, however, that the analysis is not straightforward as the right-hand side of the dynamic equation is discontinuous. In fact, considering the form of $Q_w$ in \eqref{ctrl:ChaplyginOLwMat}, the closed-loop dynamics are not defined at $w_1=0$. This is even more troublesome when we consider that we wish to regulate the system to a configuration satisfying $w_1=0$.

With this in mind, we will determine the asymptotic behaviour of the system in two steps: Firstly, it is shown that the choice of $D_i$ in \eqref{ctrl:CtrlLaw} has the consequence that, provided that $w_1(0)\neq 0$, then $w_1(t)$ cannot reach zero in finite time. This means that the closed-loop dynamics are well defined for all finite time. The second step is to show that the system cannot be positively invariant on the set $U$. As a consequence, we show that $w_1$ tends towards zero asymptotically but will not reach this configuration in finite time.

Our result requires the following Lemma: 
\begin{lem}\label{stab:fLem}
	Any real valued function $f(x)$ satisfies the inequality,
	\begin{equation}
		-\frac{1}{x_2-x_1}\left(\int_{x_1}^{x_2}f(x)dx\right)^2
		\geq
		-\int_{x_1}^{x_2}f^2(x)dx
	\end{equation}
	where $x_2 > x_1$ are in the domain of $f$.
\end{lem}
\begin{pf}
The proof is provided in the appendix. \qed
\end{pf}

It will now be shown that any solution to the closed-loop dynamics \eqref{ctrl:ChaplyginCLw} cannot satisfy $w_1(T) = 0$ for any finite time $T<\infty$.
\begin{lem}\label{stab:z1neq0}
	The set $U$ is positively invariant. That is, if $(w(0),p(0))\in U$, $(w(t),p(t))\in U$ for all time $t\geq 0$.
\end{lem}
\begin{pf}
	First note that the time derivative of $\mathcal{H}_d$ satisfies
	\begin{equation}\label{stab:HdDer}
		\begin{split}
			\dot{\mathcal{H}}_d
			&=
			-\nabla_p^\top\mathcal{H}_dD_d\nabla_p\mathcal{H}_d \\
			&<
			\nabla_p^\top\mathcal{H}_dD_i\nabla_p\mathcal{H}_d \\
			&\leq
			0.
		\end{split}
	\end{equation}
	As $\mathcal{H}_d$ is quadratic in $p,w$, \eqref{stab:HdDer} this implies that for any solution with initial conditions in $U$, $p(t)$ and $w(t)$ will be bounded over any time interval in which the solution is contained within $U$. We denote such an interval as $\Delta t = [0,T)$. Considering $Q_w$ in \eqref{ctrl:ChaplyginOLwMat}, as $p(t)$ is bounded $\Delta t$, $\dot{w}_1(t)$ is bounded on the same time interval. Boundedness of $\dot w_1$ implies that $\lim_{t\to T}w_1(t)$ exists for all $T$.

	\sloppy Now, for the sake of contradiction, assume that $\lim_{t\to T}w_1(t) = 0$ for some finite $T\in [t_0,\infty)$. Taking any interval $[t_1,T]$, such that $t_1 \geq 0$, pick $t'$ such that $w_1(t') = \max\{w_1(t)\} \ \forall t\in[t_1,T]$. From \eqref{stab:HdDer} it can be verified that time derivative of $\mathcal{H}_d$ satisfies
	\begin{equation}
		\begin{split}
			\dot{\mathcal{H}}_{d}(t) 
			\leq 
			-\frac{k}{w_1^2(t)}\nabla_{p_2}\mathcal{H}_{d}^2(t)
			=
			-\frac{k}{w_1^2(t)}\dot{w}_1^2(t)
		\end{split}
	\end{equation}
	Integrating with respect to time from $t'$ to $T$
	\begin{equation}
		\begin{split}
			\mathcal{H}_{d}(T) - \mathcal{H}_{d}(t')
			&\leq 
			-\int_{t'}^{T}\frac{k}{w_1^2(t)}\dot{w}_1^2(t) dt
		\end{split}
	\end{equation}
	As $w_1(t') = \max\{w_1(t)\}\forall t\in[t_0,T]$,
	\begin{equation}
		\begin{split}
			\mathcal{H}_{d}(T) - \mathcal{H}_{d}(t')
			&\leq 
			-\frac{k}{w_1^2(t')}\int_{t'}^{T}\dot{w}_1^2(t) dt
		\end{split}
	\end{equation}
	Applying Lemma \ref{stab:fLem} to this inequality yields
	\begin{equation}
		\begin{split}
			\mathcal{H}_{d}(T) - \mathcal{H}_{d}(t')
			&\leq 
			-\frac{k}{w_1^2(t')}\frac{1}{T-t'}\left(\int_{t'}^{T}\dot{w}_1(t) dt\right)^2 \\
			&\leq 
			-\frac{k}{w_1^2(t')}\frac{1}{T-t'}\left(w_1(T) - w_1(t')\right)^2 \\
			&\leq 
			-\frac{k}{w_1^2(t')}\frac{1}{T-t'}w_1^2(t') \\
			&\leq 
			-\frac{k}{T-t'}.
		\end{split}
	\end{equation}
	As $T - t' \leq T - t_1$ is arbitrarily small, the right hand side of this inequality can be made arbitrarily large by choosing a small enough time interval. However, $\mathcal{H}_d$ is lower bounded, thus we have a contradiction. Thus, we conclude that there is no finite $T$ such that $\lim_{t\to T}w_1(t) = 0$. As a consequence, $U$ is positively invariant.
	\qed
\end{pf}

As the set $U$ is positively invariant, the closed-loop dynamics \eqref{ctrl:ChaplyginCLw} are well defined for all time. We now show that $w_1(t)$ tends to zero asymptotically. This will be done by considering two properties. Firstly, as $\dot{\mathcal{H}}_d\leq 0$, the trajectories $(w(t),p(t))$ are confined to a compact set. Secondly, it is shown that there is no subset of $U$ that satisfies $\dot{\mathcal{H}}_d = 0$ identically. Combining these two properties, it can be deduced that $w_1\to 0$.
\begin{lem}\label{NoInvarSet}
	Consider the closed-loop dynamics \eqref{ctrl:ChaplyginCLw}. On the set $U$ there is no solution to $(w(t),p(t))$ satisfying $\dot{\mathcal{H}}_d = 0$ identically.
\end{lem}

\begin{pf}
	From \eqref{stab:HdDer}, it can be seen that time derivative of $\mathcal{H}_{d}$ satisfies
	\begin{equation}\label{HamDeriv}
		\dot{\mathcal{H}}_{d}
		\leq
		-p^\top M^{-1}\hat DM^{-1}p.
	\end{equation}	
	As $\hat D, M > 0$, for \eqref{HamDeriv} to be identically equal to zero, $p$ must be identically equal to zero. This means that $\dot{p} = 0$ along such a solution.
	
	Evaluating the $\dot{p}$ dynamics of \eqref{ctrl:ChaplyginCLw} at $p = \dot{p} = 0$ results in
	\begin{equation}\label{stab:NoInvarSet1}
		\begin{split}
			-Q_w^\top(w)Lw
			=
			-\left[Q^\top(z)\frac{\partial^\top f_w}{\partial z}\right]\bigg|_{z = f_w^{-1}(w)} Lw
			=
			0_2.
		\end{split}
	\end{equation}
	Recalling that $Q_z$ has a left annihilator given by \eqref{ctrl:QzPerp}, \eqref{stab:NoInvarSet1} is satisfied if
	\begin{equation}\label{stab:NoInvarSet2}
		\frac{\partial^\top f_w}{\partial z}\bigg|_{z = f_w^{-1}(w)} Lw = Q_z^\perp\big|_{z = f_w^{-1}(w)}a(w),
	\end{equation}
	where $Q_z^\perp$ is defined by \eqref{stab:NoInvarSet1} and $a\in\mathbb{R}$ is an unknown, possibly state dependant, function.
	Rearranging \eqref{stab:NoInvarSet2} results in
	\begin{equation}\label{stab:NoInvarSet3}
		\begin{split}
			Lw &= \frac{\partial^{\top} f_w^{-1}}{\partial w}Q_z^\perp\big|_{z = f_w^{-1}(w)} a.
		\end{split}
	\end{equation}
	Using the definition of $f_w^{-1}$ in \eqref{control:wTransformInv} and $Q_z^\perp$ in \eqref{ctrl:QzPerp}, \eqref{stab:NoInvarSet3} can be evaluated to find
	\begin{equation}\label{stab:NoInvarSet4}
		L
		\begin{bmatrix}
			w_1 \\ w_2 \\ w_3
		\end{bmatrix}
		=
		\begin{bmatrix}
			1 & w_2+w_3 & w_1w_3 \\
			0 & w_1 & 0 \\
			0 & w_1 & \frac12 w_1^2
		\end{bmatrix}
		\begin{bmatrix}
			-w_1w_2-w_1w_3 \\ 0 \\ 1
		\end{bmatrix}
		a.
	\end{equation}
	The second row of \eqref{stab:NoInvarSet4} implies that $w_2 = 0$. Substituting $w_2$ into the first row of \eqref{stab:NoInvarSet4} implies that $w_1 = 0$. However, such a solution is not contained in $U$, thus, there is no trajectory in $U$ such that $\dot{\mathcal{H}}_d = 0$ identically. \qed
\end{pf}

We are now in a position to determine the asymptotic behaviour of the closed-loop system \eqref{ctrl:ChaplyginCLw}. The typical approach to verifying asymptotic properties of pH systems is to first show that the system is stable as $\mathcal{H}_d(t) \leq \mathcal{H}_d(0)$. Then asymptotic stability is shown by application of LaSalle's theorem together with some detectability requirements. Here, there are two problems with this approach. Firstly, the dynamics are ill-defined at $w_1 = 0$, thus, the point $(w,p)=(0_{3\times 1},0_{2\times 1})$ cannot be an equilibrium---although it behaves just like one in the sense that if $(w,p)$ starts small, it stays small. Secondly, as the system dynamics are not defined for $w_1=0$, LaSalle's theorem does not apply. The following Proposition provides an argument which is similar in nature to LaSalle's theorem to show that $w_1(t)$ tends towards $0$. Considering the transformation \eqref{control:wTransformInv}, this means that $z(t)$ tends towards $0_{3\times 1}$, satisfying the control objective.

\begin{prop}\label{stab:SaympStab}
	Consider the closed-loop dynamics \eqref{ctrl:ChaplyginCLw} with initial conditions $(w(0),p(0))\in U$. The system verifies:
	\begin{enumerate}[(i)]
	\item For each $\epsilon > 0$ there exists a $\delta(\epsilon) > 0$ such that $||(w(0),p(0))||<\delta\implies ||(w(t),p(t))||<\epsilon$.
	\item $\lim_{t\to\infty}q(t) = 0_{3\times 1}$.
	\end{enumerate}
\end{prop}
\begin{pf}
	For this proof, we let $x = (p,w)$.
	Noting that as $\dot{\mathcal{H}}_d \leq 0$ for all time, $\mathcal{H}_d(t) \leq \mathcal{H}_d(0)$. As $\mathcal{H}_d$ is quadratic in $w$ and $p$, claim (i) can be verified to be true. Furthermore, this means that the set
	\begin{equation}
		\{x|\mathcal{H}_d(x)\leq\mathcal{H}_d(0)\}
	\end{equation}
	is both bounded and positively invariant.

	The proof of claim (ii) follows from similar argument to LaSalle's invariance principle. The proof is as follows:
	
	First note that $\lim_{t\to\infty}\mathcal{H}_{d} = \mathcal{H}_L$ exists and is in the set $[0,\mathcal{H}_{d}(0)]$ as $\mathcal{H}_{d}(t)$ is monotonic and bounded below by zero.
	Now define the set
	\begin{equation}
		V = U\cap \{x|\mathcal{H}_d(x)\leq\mathcal{H}_d(0)\}
	\end{equation}
	which is bounded. By Lemma \ref{stab:z1neq0}, together with claim (i), the set $V$ is positively invariant.
	Let $x(x_0,t)$ denote the solution such the $x(x_0,0) = x_0 \in V$. 
	
	Consider a solution $x(x_0,t)$ to the system. As the right hand side of \eqref{ctrl:ChaplyginCLw} is smooth on $V$, it is locally Lipschitz. Thus, the solution $x(x_0,t)$ exists and is unique for all time.
	
	By the Bolzano-Weierstrass theorem, the solution admits an accumulation point as $t\to\infty$. The set of all accumulation points is denoted $L^+$. Furthermore, $L^+$ is compact and $x(t)\to L^+$ as $t\to\infty$. (See Section C.3 Khalil for details).
	
	Now suppose that $W = L^+\cap V\neq\emptyset$. By definition, for each $y\in W$, there exists a sequence $t_n$ such that $\lim_{n\to \infty}x(t_n) = y$. As $\mathcal{H}_d$ is continuous and $\lim_{t\to\infty}\mathcal{H}_d = \mathcal{H}_L$, $\mathcal{H}_d(W) = \mathcal{H}_L$.
	
	By the continuity of solutions on $V$ and claim (i), a solution $x(y,t)$ is contained in $W$. Thus, such a solution satisfies $\dot{\mathcal{H}}_d(t) = 0$.
		
	But by Proposition \ref{NoInvarSet}, there is no solution in the set $U$ satisfying $\dot{\mathcal{H}} = 0$ identically. Thus we conclude that $W=\emptyset$ $L^+$ is contained in the set 
	\begin{equation}
		\bar{V}\setminus V = \{x|\mathcal{H}(x)\leq\mathcal{H}(0),w_1 = 0\}.
	\end{equation}
	As $x(t)\to L^+$, $w_1\to 0$.
	
	Considering the transformation $f_w^{-1}$ in \eqref{control:wTransformInv} and the fact that $w(t)$ is bounded, $w_1(t)\to 0$ implies that $z(t)\to 0_{3\times 1}$. Then, considering the transformation $f_z$ in \eqref{control:zTransform}, $z(t)\to 0_{3\times 1}$ implies that $q(t)\to 0_{3\times 1}$ as desired. \qed
\end{pf}

Notice that although $q$ tends towards the origin, the asymptotic behaviour of $p$ has not been established. Clearly $p(t)\in\mathcal{L}_\infty$ as $\frac{1}{2}p^\top M^{-1}p < \mathcal{H}_d(t) \leq \mathcal{H}_d(0)$ for all time. Further analysis is considered beyond the scope of this paper and left as future work. 


\section{Simulation Results}\label{simulation}

In order to demonstrate the effectiveness of the control strategy, a numerical simulation was preformed. The parameters used for the open-loop Chaplygin system were $m=2, J=1, l=1$ and
\begin{equation}
	\begin{split}
		D
		&=
		\begin{bmatrix}
			\frac{1}{\sqrt{0.1+p_1^2}} & 0 \\
			0 & \frac{1}{\sqrt{0.1+p_2^2}}
		\end{bmatrix}.
	\end{split}
\end{equation}
The expression $\frac{1}{\sqrt{0.1+p_i^2}}$ is an approximation of Coulomb friction \citep{Gomez-Estern2004} and assumed to be unknown for control purposes. The inertial parameters, $m$ and $J$, are also assumed to be unknown.

The control law \eqref{ctrl:CtrlLaw} was utilised for control with the following parameters:
\begin{equation}
	\begin{split}
		L&=\operatorname{diag}(2, 0.5, 0.8)\\
		k&=0.1\\
		\hat D&=\operatorname{diag}(4,8).
	\end{split}
\end{equation}

\begin{figure}
	\centering
	\includegraphics[width=0.5\textwidth]{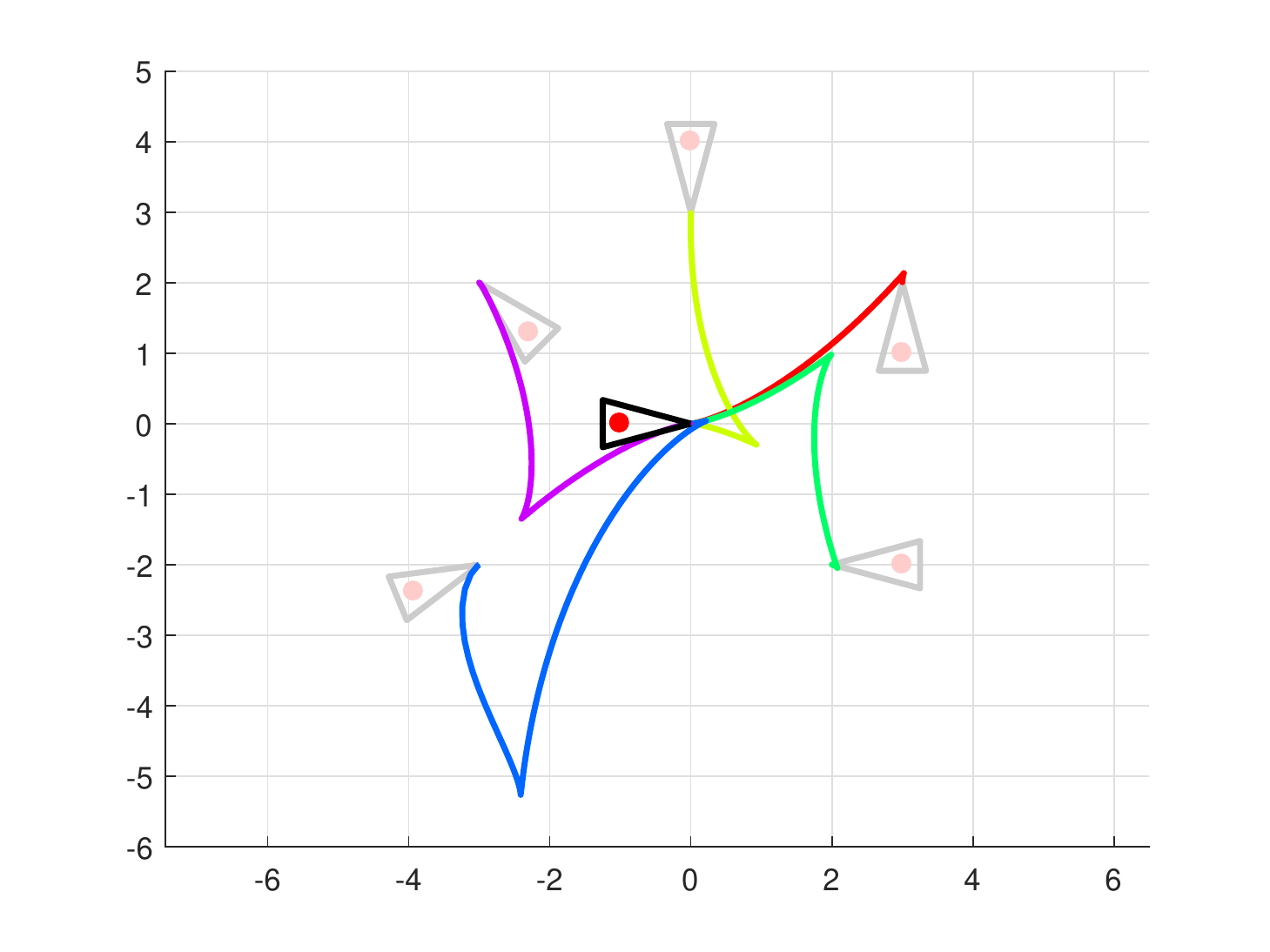}
	\caption{Chaplygin sleigh initialised from several initial conditions. The system is controlled by a discontinuous, potential energy shaping controller which drives the configuration to the origin.}
	\label{simulation:Sim1Path}
\end{figure}

\begin{figure}
	\centering
	\includegraphics[width=0.5\textwidth]{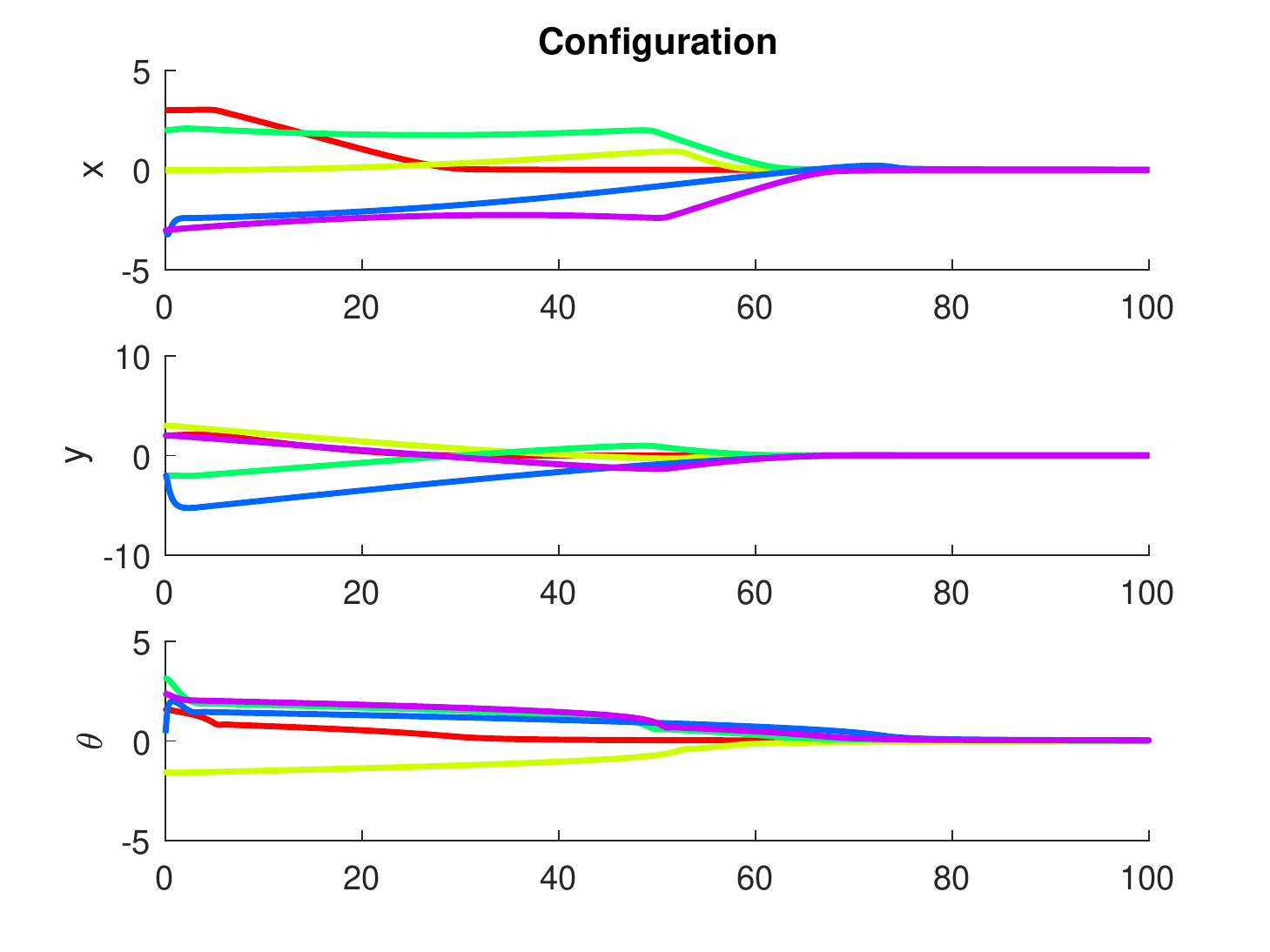}
	\caption{Time history of the configuration variables of the closed-loop Chaplygin sleigh system from several simulation scenarios.}
	\label{simulation:config}
\end{figure}

\begin{figure}
	\centering
	\includegraphics[width=0.5\textwidth]{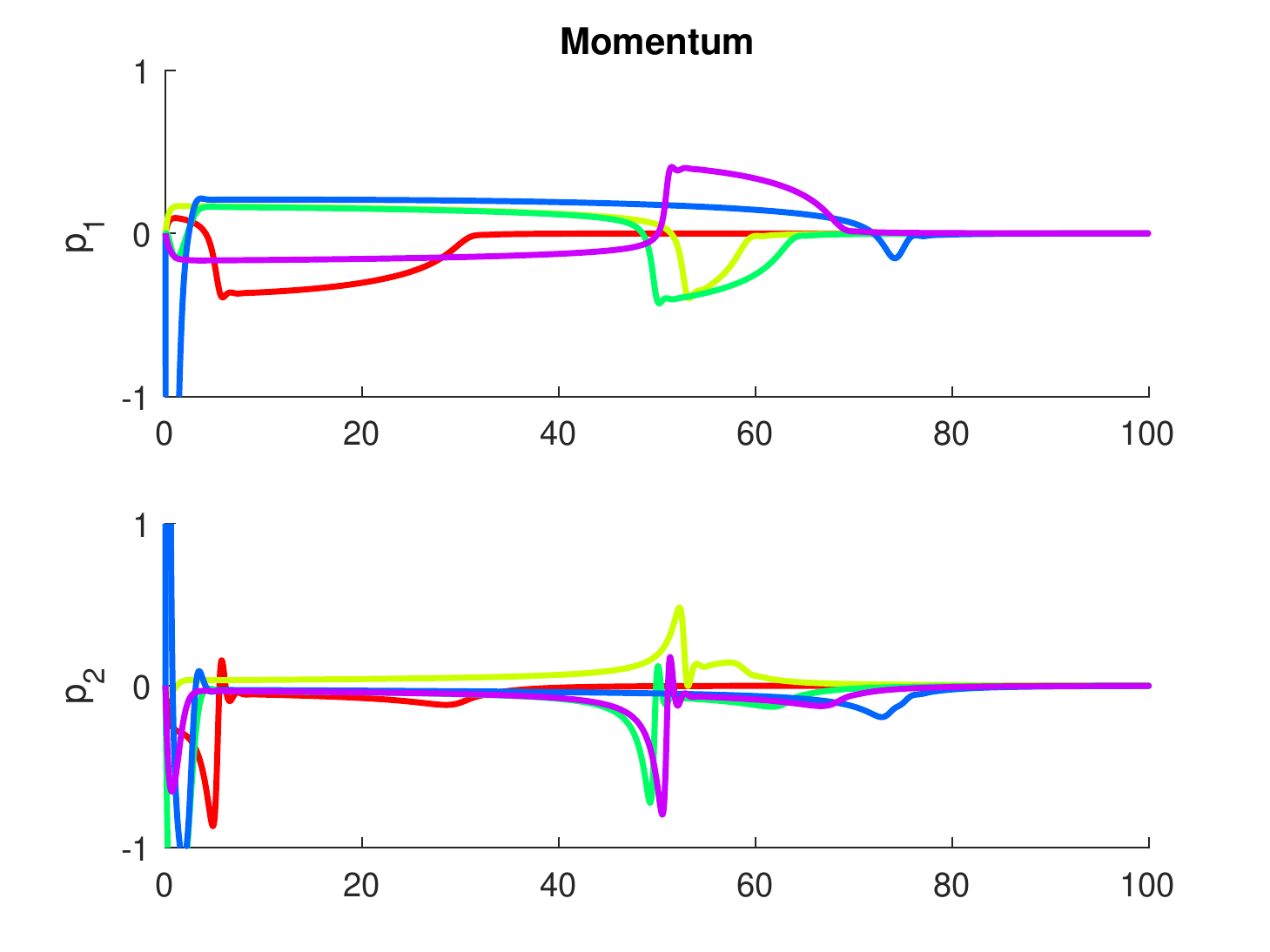}
	\caption{Time history of the momentum variables of the closed-loop Chaplygin sleigh system from several simulation scenarios.}
	\label{simulation:momentum}
\end{figure}

\begin{figure}
	\centering
	\includegraphics[width=0.5\textwidth]{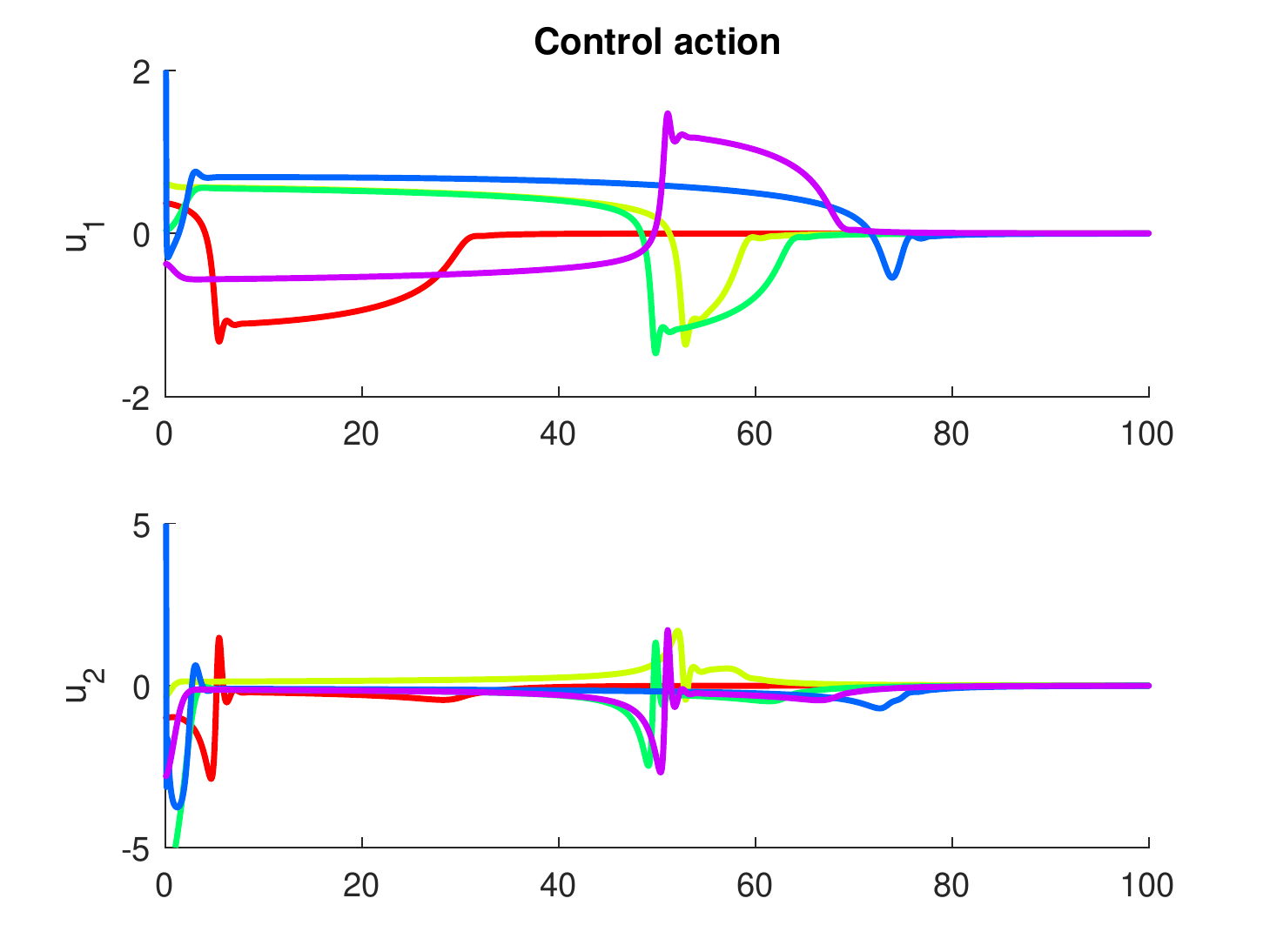}
	\caption{Time history of the control signals of the closed-loop Chaplygin sleigh system from several simulation scenarios.}
	\label{simulation:control}
\end{figure}

The Chaplygin sleigh was initialised from an assortment of positions and the simulation was run for 100 seconds. The resulting path of each simulation is shown in Figure \ref{simulation:Sim1Path} where the ghosted images of the sleigh represent the initial positions and the solid image of the sleigh is the target final position. The time histories of the configuration, momentum and control signals for each run can be found in Figures \ref{simulation:config}, \ref{simulation:momentum} and \ref{simulation:control}, respectively. Notice that although we have not proved convergence of the functions $p(t)$ or $u(t)$, they appear well behaved in the numerical simulation. 

Notice that the simulation plotted in blue, which was initialised at $(x,y,\theta)=(-3, -2, \frac18 \pi)$, takes a rather inefficient path to the origin. This is because it was initialised close to the set parametrised by $w_1=z_1=\theta =0$, on which the control law and closed-loop are not defined. As a result of this singularity, initial conditions close to $\theta=0$ will have a large initial closed-loop Hamiltonian $\mathcal{H}_d$. To dissipate this energy from the closed-loop, the system traverses a long path before converging. Simulations with initial conditions away from this singularity take more `natural' paths.



\section{Conclusion}\label{conclusion}

In this paper we presented a discontinuous control law for the Chaplygin sleigh system which is robust against both the inertial parameters and damping structure of the open-loop system. The proposed control law is successful in driving the configuration of the system to the origin. The results were demonstrated by performing numerical simulations to verify the theoretical claims. In future work, we aim to both extend the analysis to characterise the behaviour of the momentum and control signals as well as extend the controller to apply to a wider class of nonholonomic systems.

\appendix
\section{Proofs}
\begin{pf*}{Proof of Lemma \ref{stab:fLem}}
	By the Schwarz inequality \citep{lieb2001analysis}, any two real valued functions $f(x)$, $g(x)$ satisfy
	\begin{equation}\label{schwartz}
		\left(\int_{x_1}^{x_2}f(x)g(x)dx\right)^2
		\leq
		\int_{x_1}^{x_2}f^2(x)dx
		\int_{x_1}^{x_2}g^2(x)dx.
	\end{equation}
	Taking $g(x) = 1$, \eqref{schwartz} simplifies to
	\begin{equation}
		\begin{split}
			\left(\int_{x_1}^{x_2}f(x)dx\right)^2
			&\leq
			\int_{x_1}^{x_2}f^2(x)dx
			\int_{x_1}^{x_2}1dx \\
			&\leq
			(x_2-x_1)\int_{x_1}^{x_2}f^2(x)dx \\
			\frac{1}{x_2-x_1}\left(\int_{x_1}^{x_2}f(x)dx\right)^2
			&\leq
			\int_{x_1}^{x_2}f^2(x)dx \\
		\end{split}
	\end{equation}
	Taking the negative of this inequality results in
	\begin{equation}
		-\frac{1}{x_2-x_1}\left(\int_{x_1}^{x_2}f(x)dx\right)^2
		\geq
		-\int_{x_1}^{x_2}f^2(x)dx \\
	\end{equation}
	as desired. \qed
\end{pf*}
\end{document}